\documentclass[twocolumn ,showpacs,floatfix,preprintnumbers,prb]{revtex4-1}
\usepackage{graphicx}
\usepackage{float}
\usepackage{dcolumn}
\usepackage{bm}
\usepackage{amssymb}

\begin{document}

\title{Spectral, optical and transport properties of the adiabatic anisotropic Holstein model: Application to slightly doped organic semiconductors}

\author{C.A. Perroni, A. Nocera, V. Marigliano Ramaglia, and V. Cataudella  }
\affiliation{CNR-SPIN and Dipartimento di Scienze Fisiche, Univ. di Napoli ``Federico
II'', I-80126 Italy}

\email{perroni@na.infn.it}

\begin{abstract}
Spectral, optical and transport properties of an anisotropic three-dimensional Holstein model are studied within the adiabatic approximation.
The parameter regime is appropriate for organic semiconductors used in single crystal based field effect transistors.
Different approaches have been used to solve the model: self-consistent Born approximation valid for weak electron-phonon coupling,
coherent potential approximation  exact for infinite dimensions, and numerical diagonalization for finite lattices. With increasing temperature, the width of the spectral functions gets larger and larger making the approximation of quasi-particle less accurate. On the contrary, their peak positions are never
strongly renormalized in comparison with the bare ones. As expected, the density of states is characterized by an exponential tail corresponding
to localized states at low temperature. For weak electron-lattice coupling, the optical conductivity follows a Drude behavior, while, for intermediate
electron-lattice coupling, a temperature dependent peak is present at low frequency. For high temperatures and low particle densities, the mobility always
exhibits a power-law behavior as function of temperature. With decreasing the particle density,  at low temperature, the mobility shows a transition from metallic to insulating behavior. Results are discussed in connection with available experimental data.
\end{abstract}
\maketitle

\section{introduction}
Organic field-effect transistors (OFETs) play a key role in the field of plastic electronics. Recently,  single crystal OFETs have been developed showing charge mobility at least one order of magnitude larger than that of thin films. \cite{hasegawa} The most promising are those based on polyacenes, such as pentacene and rubrene, that exhibit a strong anysotropy and the largest mobility measured in organic semiconductors. \cite{morpurgo}

In spite of many applications based on such devices, the intrinsic transport mechanism acting in high mobility OFETs is not fully understood. At low temperatures, an insulating behavior takes place with a sharp fall of resistivity. At temperatures higher than $100 K$, the mobility $\mu_p$
of these systems exhibits a power-law behavior ($\mu_p \sim T^{-\delta}$), \cite{morpurgo,ostroverkhova,nature} which cannot be simply ascribed
to band transport. \cite{cheng} Moreover, in some systems, starting from room temperature, there is a crossover from band-like to activated hopping behavior. \cite{corop,cheng1}
Very recently, in order to better investigate the properties of the quasi-particles involved in the transport mechanism, many experiments based on spectroscopic probes such as photoemission \cite{arpes,arpes1,arpes2,kakuta} and optical conductivity \cite{li,petrenko,fisher} have been performed in pentacene and rubrene. Photoemission experiments suggest that the quasi-particle energy dispersion does not exhibit a strong mass renormalization even if the width of the peaks increases significantly with temperature. Optical conductivity measurements cannot be easily explained in terms of a Drude-like response, since they show a low frequency peak whose origin is not clear. Both kinds of experiments indicate moderate values of electron-phonon (el-ph) coupling pointing out that polaron formation should not play a prominent role for the interpretation of data.

From the theoretical point of view, the calculated scattering lengths of the charge carriers are too short to be compatible with the band-like transport mechanism found at intermediate temperatures. \cite{cheng} Therefore, in order to analyze the tunneling-hopping crossover in some transport data, a mechanism based on polaron band narrowing with increasing temperature has been proposed. \cite{hannewald2} For the theoretical analysis, the simplest model for molecular crystals, the Holstein model, \cite{holstein1,holstein2} has been quite extensively studied within several el-ph coupling regimes.
Recently, in order to take into account the anisotropy of the electronic band and the most important couplings coming from the interaction of charge carriers with low-frequency phonons, a three-dimensional anisotropic Holstein model for a single electron has been analyzed. \cite{hannewald} However, a study of
spectral and optical properties in the regime appropriate to organic semiconductors has not been performed up to now.

For high mobility systems, a very interesting and simple one-dimensional model has been recently introduced by Troisi and Orlandi,\cite{troisi_prl} where the charge carrier interacts with low-frequency modes leading to a modulation of the charge carrier hopping. The model is proposed for an approximate description of rubrene and it is somehow very close to the Su-Schrieffer-Heeger (SSH) model introduced in a different context.\cite{SSH} The model has been studied within the adiabatic limit (phonons obey a classical dynamics) by using an approximated dynamical approach. Some results have been confirmed and extended by Ciuchi and Fratini, \cite{fc} who, instead, used a static approach, which is equivalent to the classical problem of a quantum particle in the presence of disorder due to electron-lattice coupling. \cite{mahan} The main results of these studies is that a power law for the mobility temperature dependence can be recovered within the proposed model and that the charge carriers involved in the transport undergo a "dynamical localization" with increasing temperature. Very recently, some of us have made a systematic study of the transport properties of the SSH model in the adiabatic limit by using the Kubo formula for the conductivity showing that the inclusion of the vertex renormalizations is relevant for the interpretation of experimental results. \cite{vittoriocheck} Moreover, in this paper, spectral and optical properties of finite density systems have been discussed in connection with experimental data.

It is clear that one-dimensional adiabatic models suffer of severe limitations, of which the main is that electronic states are always localized. \cite{anderson} Moreover, ingredients such as band anisotropy, small but finite carrier density are necessary for a correct description of the systems. Therefore, in this paper, we study the Holstein model proposed in Ref.[\onlinecite{hannewald}] within the adiabatic approach focusing on the weak to
intermediate el-ph coupling regime which is relevant for high mobility organic semiconductors. Different methods have been used to solve the model: self-consistent Born approximation (SCBA) accurate for weak el-ph coupling, coherent potential approximation (CPA) exact for infinite dimensions, and exact diagonalization (ED) for finite lattices.

The calculated quantities are the spectral, optical and transport properties of the model.
With increasing temperature, the width of the spectral functions gets larger and larger making the quasi-particles less defined.
On the contrary, the spectral functions show peaks which are weakly renormalized in comparison with
those of the bare bands. The marked width of the spectral functions gives rise to densities of states with a low energy exponential tail
increasing with temperature. At low temperatures, this tail corresponds to localized states and gives rough indications for the energy position of the mobility edge. With increasing temperature, in the regime of low carrier doping investigated in this paper, the chemical potential always enters the energy region of the tail. These features allow to reconcile the results provided by the Angle Resolved Photoemission Spectroscopy (ARPES) data (band-like description) with the trend that charge carriers appear more localized at high temperature.

The crossover between low to high temperature behavior considered in the spectral properties has a relevant effect also on the optical and transport properties. For weak el-ph coupling, the optical conductivity follows a Drude behavior and the mobility fully described by SCBA can be ascribed to band transport. In the intermediate el-ph coupling regime, vertex corrections become important and
an exact diagonalization is needed. In this regime, at low temperature, a temperature dependent low frequency peak appears in the conductivity,
while, at higher temperatures, the spectral features broaden over quite a large frequency range. The presence of this peak confirms the
findings of a previous paper, \cite{vittoriocheck} and compares well with recent experimental results.\cite{li,petrenko,fisher} Finally, the mobility exhibits
a power-law behavior for any high temperature. For very small particle density, at low temperature, the mobility shows a transition from metallic to
insulating behavior with decreasing the particle density. Results are discussed in connection with available experimental data.

The paper is organized as follows. In section II the model and the methods of solution are presented, in section III spectral properties are analyzed, in section IV optical and transport properties are discussed.

\section{The model}

The spectral, optical and transport properties will be studied within the Holstein model\cite{holstein1,holstein2} in the adiabatic limit including the anisotropy important for the description of organic semiconductors. \cite{hannewald} It can be summarized in the following model hamiltonian:
\begin{equation}
H= \sum_{\vec{R}_i} \frac{{p}^2_{\vec{R}_i}}{2 m}+\sum_{\vec{R}_i} \frac{ k x_{\vec{R}_i}^{2}}{2}
+H_{el},
\label{h}
\end{equation}
where $x_{\vec{R}_i}$ and ${p}_{\vec{R}_i}$ are the classical oscillator displacement and momentum, respectively, relative to the site $\vec{R}_i$ of a cubic lattice with parameter $a$, $m$ the oscillator mass, $k$ the elastic constant, and the electronic part $H_{el}$ is
\begin{equation}
H_{el}=- \sum_{\vec{R}_i, \vec{\delta}} t_{|\vec{\delta}|} c_{\vec{R}_i}^{\dagger}c_{\vec{R}_i+\vec{\delta}} +
\alpha \sum_{\vec{R}_i} x_{\vec{R}_i} c_{\vec{R}_i}^{\dagger} c_{\vec{R}_i}  .
\label{hel}
\end{equation}
In eq. (\ref{hel}), $t_{|\vec{\delta}|}$ is the bare electron hopping toward the nearest neighbors $\vec{\delta}$, $c_{\vec{R}_i}^{\dagger}$ and
$c_{\vec{R}_i}$ are the charge carrier creation and annihilation operators, respectively, and
$\alpha$ is the coupling constant that controls the link between the electron density and lattice displacement. We use units such that lattice parameter $a=1$, Planck constant $\hbar=1$, Boltzmann constant $k_B=1$, and electron charge $e=1$.

The treatment is semiclassical since the electron dynamics is fully quantum, while the ion dynamics is assumed classical.
The latter approximation is well justified from the typical values of phonon frequencies $\omega_0 =\sqrt{k/m}$ and hopping constants of high mobility organic semiconductors such as pentacene or rubrene. Following Ref.[\onlinecite{hannewald}],
$\omega_0 \simeq 10 meV$, $t_z \simeq 100 meV$, $t_x \simeq 50 meV$, $t_y \simeq 20 meV$ leading to an adiabatic ratio
$\gamma=\omega_0/t_z = 0.1$. Therefore, all the results reported in this paper are valid for temperatures $T \geq \omega_0 \simeq 120 K $.

In the adiabatic limit, the following quantity $\lambda$
\begin{equation}
\lambda=\frac{\alpha^2}{4 k t_z} \label{lamb}
\end{equation}
correctly describes the strength of the el-ph coupling. Actually, in the adiabatic limit, the calculation is equivalent to the
classical problem of quantum particles in the presence of diagonal disorder $\epsilon_{\vec{R}_i}=\alpha x_{\vec{R}_i}$ due to electron-lattice coupling. The disorder is distributed according to the probability function of the lattice displacements $P \left( \{ x_{\vec{R}_i} \} \right)$, that has to be self-consistently calculated as a function of electron-phonon coupling $\lambda$, temperature $T$ and particle density $c=N_p/L$, with $N_p$ number of particles and $L$ number of sites of the cubic lattice.

In this paper, we will focus on the weak to intermediate electron-lattice regime that seems to be appropriate to systems like high mobility organic semiconductors. Simulations will concentrate on temperatures lower than the typical electron energy of the order
of $t_z$  (this temperature range is extensively investigated in many OFETs), and they will not explore the high temperature regime.
\cite{millis,gunnarsson} Moreover, in most OFETs, the induced doping is not very high (typically much smaller than one charge carrier for ten molecules),
therefore, in the following, we will investigate the regime of low doping (up to $c=0.01$). For this regime of parameters, the probability function
of the lattice displacements $P \left( \{ x_{\vec{R}_i} \} \right)$ shows very tiny deviations from the distribution of free oscillators
(it has been checked also in a recent paper \cite{vittoriocheck} for a related model), so that one can correctly assume
\begin{equation}
P \left( \{ x_{\vec{R}_i} \} \right)= \left( \frac{2m\pi} {\beta} \right)^{L/2}
\exp\left[-\beta\frac{k}{2}\sum_{i}x_{\vec{R}_i}^{2}\right],
\label{distri}
\end{equation}
with $\beta=1/T$. The mean value of an observable $O \left( \{ x_{\vec{R}_i} \} \right)$ over the distribution $P \left( \{ x_{\vec{R}_i} \} \right)$
is indicated by the standard symbol
\begin{equation}
\left\langle O \left( \{ x_{\vec{R}_i} \} \right) \right\rangle=\int  \left( \prod_{\vec{R}_i} d x_{\vec{R}_i} \right)    P \left( \{ x_{\vec{R}_i} \} \right)
O \left( \{ x_{\vec{R}_i} \} \right).
\label{distri}
\end{equation}

In the next three subsections, the methods of solution for the considered model are discussed.

\subsection{Self-consistent Born approximation}
This approximation is valid in the weak el-ph coupling regime. \cite{mahan} It does not rely on the entire distribution of $\epsilon_{\vec{R}_i}$, but only on
the first two moments: $<\epsilon_{\vec{R}_i}>=0$ and $<\epsilon_{\vec{R}_i} \epsilon_{\vec{R}_j}>=\delta_{\vec{R}_i,\vec{R}_j} \tau$, with $ \tau= 4 T t_z \lambda$. Actually, temperature and el-ph coupling are cooperative to increase the interaction between electron and phonons.

SCBA is an approximation for the averaged retarded Green function or equivalently for the retarded self-energy. For the considered model, the self-energy is independent of the wave vector $\vec{k}$ and it is given by
\begin{equation}
\Sigma_{SCBA}^{ret} (\omega)= \frac{\tau}{L} \sum_{\vec{k}} G^{ret} (\vec{k},\omega),
\label{self}
\end{equation}
where the Dyson equation provides the link between Green function and self-energy. Once the equation in frequency is solved, one can easily get the spectral function $A(\vec{k},\omega)$ through the following equation
\begin{equation}
A(\vec{k},\omega)= -2 \Im G^{ret} (\vec{k},\omega),
\label{specfunc}
\end{equation}
and the density of states $g (\omega)$
\begin{equation}
g (\omega)= \frac{1}{2 \pi} \frac{1}{L} \sum_{\vec{k}} A(\vec{k},\omega).
\label{density}
\end{equation}
The equation for self-consistency can be solved for large lattice sizes of the order of
$L=30^3$,  that is indistinguishable from the thermodynamic limit.

In order to satisfy Ward identities, the calculation of conductivity within SCBA has to include ladder vertex corrections into the current-current correlation function of the Kubo formula. \cite{mahan} However, for the model considered in this study, these corrections are not effective, so that the conductivity is simply given by the product of spectral functions at different energies corresponding to the bubble diagram. \cite{mahan}

The quantities calculated within SCBA are reliable for not large el-ph couplings (up to $\lambda=0.3$) and
for not too high temperatures (up to $T=0.4 t_z$).

\subsection{Coherent potential approximation}
This approximation reduces to dynamical mean field theory in the limit of low particle density. \cite{economou} Therefore, it is exact for infinite dimensions. Consequently, CPA self-energy is independent of the wave vector.

CPA does rely on the entire displacement distribution and consists of the following self-consistent equation for the self-energy
\begin{equation}
\Sigma_{CPA}^{ret} (\omega)= \left\langle \frac{ \epsilon_{\vec{R}_i}  } {1-\left( \epsilon_{\vec{R}_i}-\Sigma_{CPA}^{ret} (\omega)\right)
G^{ret}_{\vec{R}_i, \vec{R}_i} (\omega)} \right\rangle,
\label{cpaeq}
\end{equation}
where the local Green function $G^{ret}_{\vec{R}_i, \vec{R}_i} (\omega)$ is given by
\begin{equation}
G^{ret}_{\vec{R}_i, \vec{R}_i} (\omega)= \frac{1}{L} \sum_{\vec{k}}
\frac{1}{ \omega+ \mu - \epsilon_{\vec{k}} - \Sigma_{CPA}^{ret} (\omega)},
\label{gcpa}
\end{equation}
with $\mu$ chemical potential and $\epsilon_{\vec{k}}$ electronic band of the model
\begin{equation}
\epsilon_{\vec{k}}=-2 t_z \cos{(k_z)}- 2 t_x \cos{(k_x)}- 2 t_y \cos{(k_y)}.
\label{bandeff}
\end{equation}
For the solution of the equations, sizes of the order of $L=26^3$ are employed, getting results very close to the thermodynamic limit.

Using CPA spectral functions, conductivity can be calculated. In the limit of infinite dimensions, there are no vertex corrections in the the current-current correlation function, so that only the bubble diagram contributes to the calculation of conductivity.\cite{economou} Therefore, while spectral properties
calculated within CPA are reliable for any strength of el-ph coupling $\lambda$ and temperature $T$, the optical and transport properties are not very accurate, especially in the regime of very low particle density.

\subsection{Exact diagonalization}
Within ED, static and dynamic quantities of the system can be exactly derived.

The electron part of the partition function can be calculated as $\left\langle Z_{el}[\{x_{\vec{R}_i}\}] \right\rangle$, where $Z_{el}[\{x_{\vec{R}_i}\}]$ is the quantum partition function of the electron subsystem given a deformation
configuration $\left\{x_{\vec{R}_i} \right\}$. This quantity is calculated through ED.
The method provides an approximation-free partition function of the model in the semiclassical limit.
The only limitation is due to the computational time being controlled by the $L\times L$ matrix diagonalization.
This constrains our analysis up to $L=12^3$. In order to reduce the size effect, periodic boundary conditions are used.

Within the same framework, it is also possible to calculate the spectral function, the density of states, and the conductivity averaging the electronic properties at a given ion displacement configuration over the entire set of configurations.\cite{dagotto} For instance, in the case of conductivity, for each configuration, one calculates

\begin{eqnarray}
&&Re[\sigma_{\rho,\rho}(\omega;\{x_{\vec{R}_i}\})]=\frac{\pi \left( 1-e^{-\beta \omega} \right)}{L \omega} \nonumber\\
&&\sum_{r \neq s} p_r \left( 1-p_s \right)\left|\left\langle r|J_{\rho}|s\right\rangle\right|^2\delta(E_r-E_s+\omega),
\label{sigma_mu}\\\nonumber
\end{eqnarray}
with $\rho=x,y,z$. In Eq. (\ref{sigma_mu}) the Fermi distribution function $p_r$
\begin{equation}
 p_r=\frac{1}{\exp[\beta (E_r-\mu)]+1}
\label{fermi}
\end{equation}
corresponds to the exact eigenvalue $E_r$, and $\left\langle r|J_{\rho}|s\right\rangle$ are the matrix elements between the exact eigenstates $|r>$ and $|s>$
of the current operator $J_{\rho}$ along the direction $\hat{e}_{\rho}$, defined as
\begin{equation}
J_{\rho} = i \sum_{\vec{R}_i, \vec{\delta} } t_{|\vec{\delta}|} \left( \vec{\delta} \cdot \hat{e}_{\rho} \right)    c_{\vec{R}_i}^{\dagger} c_{\vec{R}_i+\vec{\delta}}.
\label{current}
\end{equation}
We notice that, in contrast with spectral properties, the temperature enters the calculation not only through the displacement distribution, but also directly for each configuration through the Fermi distributions $p_r$. The numerical calculation of the conductivity is able to include the vertex corrections discarded by previous approaches.

The mobility along the direction $\rho$ is defined as
\begin{equation}
\mu_{\rho}=\lim_{\omega\rightarrow 0^+} \frac{Re[\sigma_{\rho,\rho}(\omega)]}{e c}=\lim_{\omega\rightarrow 0^+}
 \frac{\left\langle Re[\sigma_{\rho,\rho}(\omega;\{x_{\vec{R}_i}\})]  \right\rangle}{e c}.
\label{mobility_mu}
\end{equation}

In order to calculate observables for any finite lattice, we have to replace the delta function with a Lorentzian: \begin{equation}
\frac{1}{\pi}\delta(E_r-E_s+ \omega)\mapsto\frac{\eta}{(E_r-E_s+ \omega)^2+\eta^2}.
\end{equation}
This replacement does not give any problem for quantities at finite frequency, as soon as $ \omega > \eta$. In the case of mobility, this procedure has to be correctly implemented since there is the limit $\omega \rightarrow 0$. The correct expression is obtained performing the limits
in the following order: $\omega \mapsto 0$, $\eta \mapsto 0$ and $L \mapsto \infty$. This limit procedure is easily achieved
within SCBA and CPA, due to the large values of L. However, for ED, it is less trivial. A detailed analysis has allowed us
to obtain values for $\omega_{min}$ and $\eta_{min}$ that assure the determination of the correct value of the mobility for any temperature.
For the ED mobility presented in this paper, we used $ \omega_{min}=10^{-3} t_z$, $\eta_{min}=3*10^{-2} t_z$, and $L=12^3$.

In the following, we will measure energies in units of $t_z$. Next spectral, optical and transport properties are discussed.

\begin{figure}[htb]
%\centering
\flushleft
\includegraphics[width=0.55\textwidth,angle=0]{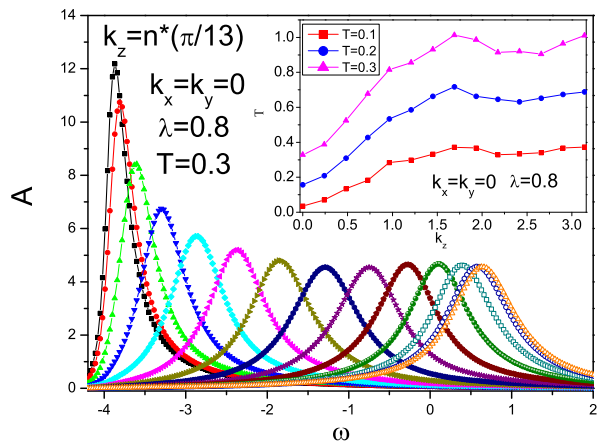}
\caption{(Color online) Spectral function within CPA at $T=0.3$ as function of frequency (in units of $t_z$) for different values of $k_z=2 \pi n / (L)^{1/3}$ (in units of $1/a$),
with $n=1,..,13$ from left to right. In the inset, the width at half height as function of $k_z$ (in units of $1/a$) for different values of temperature.
The system size is $L=26^3$. }
\label{spectral1}
\end{figure}

\section{Spectral properties}
Concerning spectral properties, focus is on the el-ph intermediate regime ($\lambda=0.8$), that is relevant
for many organic semiconductors. \cite{corop}

In Fig. \ref{spectral1}, the spectral function $A(\vec{k},\omega)$ within CPA is reported at $T=0.3$ for $k_z=2 \pi n / (L)^{1/3} $ and
$k_x=k_y=0$. However, we note that ED and CPA spectral functions are almost indistinguishable when $L$ is small.

With increasing the value of $k_z$, the shape of the spectral function changes a lot. It is more peaked at
small values of $k_z$, while it broadens at the edge zone. For this reason, we show the widths at half height in the inset of Fig. \ref{spectral1}.
The width increases at the edge zone up to a value very close to $\tau/t_z=4 \lambda T$. This quantity is related to
the width at half height of the probability distribution for the random variable $\epsilon_{\vec{R}_i}$, which gives rise to the energy fluctuations of the electron system.
Therefore, with increasing the temperature, the quasi-particles become less defined. Actually, at $T=0.3$, the behavior is more on the
border of a quasi-particle description since the width is on average nearly of the order of $t_z$. From the analysis of widths at different temperatures,
it is clear that there is a crossover from low to high $T$ behavior and the increase at a fixed point of the Brillouin zone is linear as function of temperature, in agreement with experimental data. \cite{arpes2}

\begin{figure}[htb]
\centering
\includegraphics[width=0.55\textwidth,angle=0]{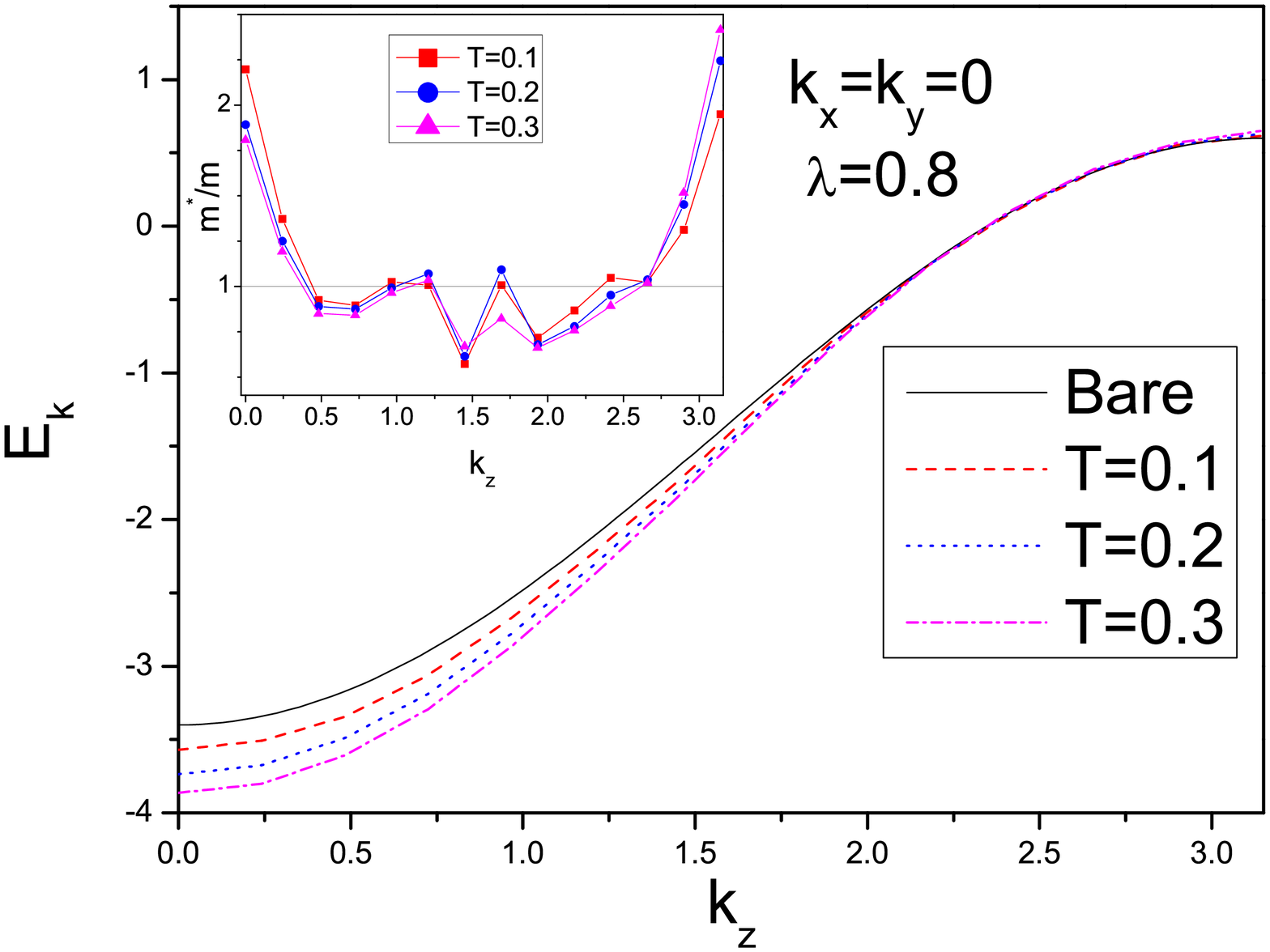}
\caption{(Color online) The quasi-particle energy within CPA as function of $k_z$ (in units of $1/a$) for different temperatures at $\lambda=0.8$. In the inset, the effective mass ratio
$m^*/m$ as function of $k_z$ (in units of $1/a$) for different temperatures at $\lambda=0.8$. The system size is $L=26^3$.}
\label{spectral2}
\end{figure}

Another interesting point is about the peak positions. In Fig. \ref{spectral2}, we report the quasi-particle energy dispersion derived by the peak positions of the spectral functions at different temperatures. Actually, the main peak of the spectral function inherits the bare band dispersion.
With increasing temperature, only the peaks at small $k_z$ show a sensible reduction, but these corresponds to narrower peaks.
Therefore, the widest peaks point towards bare band energies. This trend is also found in the effective masses. They are presented
in the inset of Fig. \ref{spectral2}. Only a factor of $2$ is gained at center and border zone. Moreover, effective mass values are quite stable
with varying the temperature. Indeed, the system is not characterized by any quasi-particle with heavy effective mass. The behavior of the quasi-particle band
and mass is different from previous proposals based on polaron mechanism. \cite{hannewald} Indeed, if polaron formation takes place, a narrowing of the quasi-particle band and a marked increase of the mass is expected with increasing temperature. Actually, this behavior is exponential in the el-ph coupling
and phonon distribution. \cite{mahan} In polaronic approaches, the entire bare band is reduced with a factor that is equal for all points of the
Brillouin zone, while in our approach there is a strong difference in the renormalization at center and border zone. Moreover, in our approach, the increase
of the width of the quasi-particle peak is playing a major role.

\begin{figure}[htb]
\centering
\includegraphics[width=0.55\textwidth,angle=0]{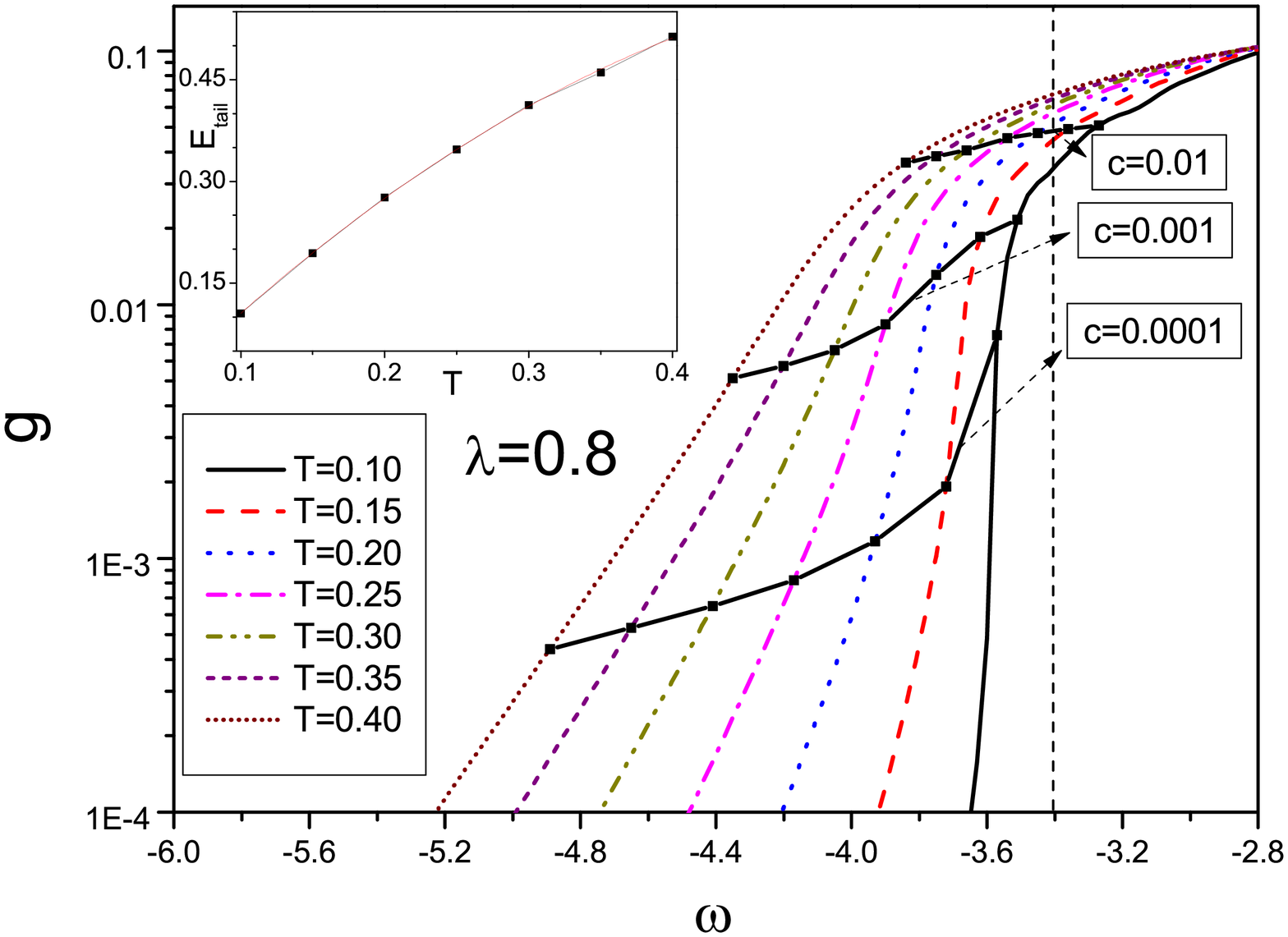}
\caption{(Color online) Density of states within CPA as function of frequency (in units of $t_z$) for different temperatures at $\lambda=0.8$.
The horizontal lines depict the chemical potential at $c=0.01$, $c=0.001$, and $c=0.0001$. The vertical dashed line at the energy $-3.4 t_z$
corresponds to the minimum of the bare band dispersion at the center zone. In the inset, the tail energy $E_{tail}$
as function of temperature at the same value of $\lambda$. The system size is $L=26^3$.}
\label{density1}
\end{figure}

These behaviors are in agreement with ARPES data, that, as mentioned in the introduction, reflect the bare electron dispersions. Only in pentacene a very weak narrowing of the band has been found. From the
analysis of the spectral functions shown in the previous figures it emerges that caution should be applied in the interpretation of the data since
width is another important parameter. It can increase with changing the system parameters making the band-like description not fully correct.

\begin{figure}[htb]
\centering
\includegraphics[width=0.55\textwidth,angle=0]{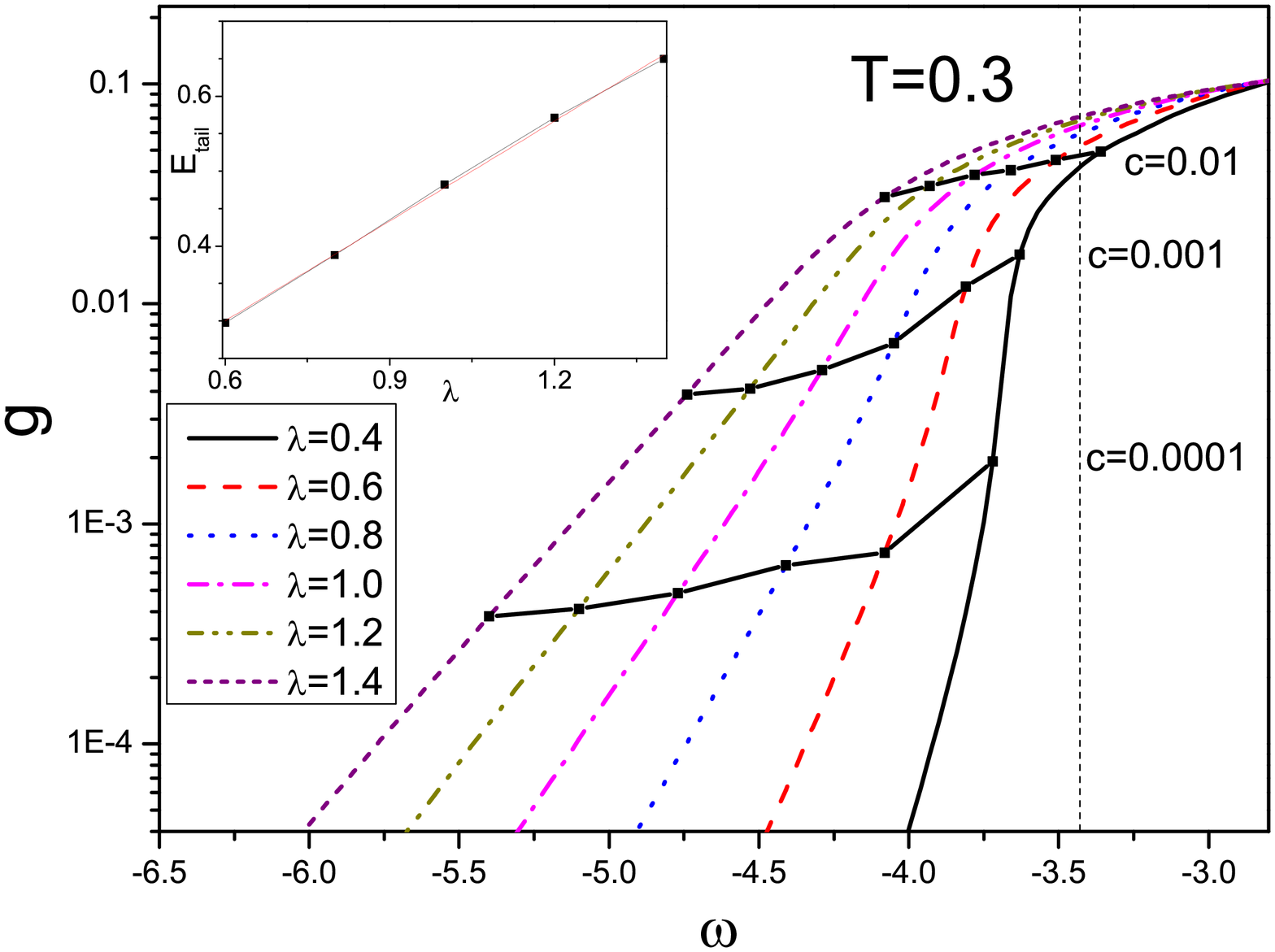}
\caption{(Color online) Density of states within CPA as function of frequency (in units of $t_z$) for different values of $\lambda$ at $T=0.3$.
The horizontal lines depict the chemical potential at $c=0.01$, $c=0.001$, and $c=0.0001$.
The vertical dashed line at the energy $-3.4 t_z$ corresponds to the minimum of the bare band dispersion at the center zone.
In the inset, the tail energy $E_{tail}$ as function of $\lambda$ at the same value of temperature. The system size is $L=26^3$.}
\label{density2}
\end{figure}

The crossover from low to high temperature behavior affects the density of states extracted from the spectral functions.
The marked width of the spectral functions gives rise to densities of states with a low energy tail.
In Fig. \ref{density1}, the density of states $g$ is shown for different temperatures at $\lambda=0.8$.
As shown in logarithmic scale, the density of states has an exponential behavior: $g(\omega) \propto e^{-\omega/E_{tail}}$.
At very low temperatures, \cite{economou} this tail is well known and it corresponds to localized states.
Moreover, it gives rough indications for the energy position of the mobility edge that should be close to the shoulder
at the beginning of the tail. We notice that the energies in the tail increase very rapidly with temperature.

In the inset of Fig. \ref{density1}, the energy $E_{tail}$ is reported as function of $T$. It shows a quadratic behavior
as function of $T$. At high temperature, it becomes of the order of half $t_z$. The temperature trend and the order of magnitude
of $E_{tail}$ represent important pieces of information since, recently, there has been an attempt to estimate the tail in the density
of states, \cite{batlogg1,batlogg2} so that this study provides some clues for the interpretation of experimental data.

It is interesting to analyze the role played by the chemical potential $\mu$ with varying the temperature. In Fig. \ref{density1}, the horizontal lines
correspond to the values of the chemical potential at $c=0.01$, $c=0.001$, and $c=0.0001$. For $c=0.01$ and low temperature, $\mu$
is outside the tail characterized by localized states. Instead, for $c=0.0001$ and low temperature, $\mu$ enters the energy tail and
will penetrate into it with increasing temperature. In any case, with increasing temperature, $\mu$ goes into the energy region of the tail.

\begin{figure}[htb]
\centering
\includegraphics[width=0.55\textwidth,angle=0]{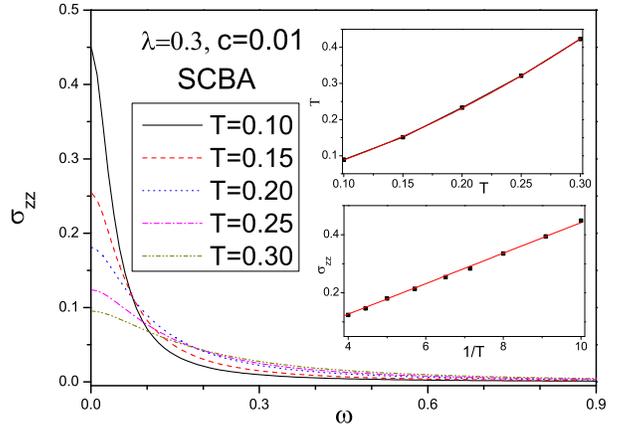}
\caption{(Color online) Optical conductivity $\sigma_{zz}$ as a function of frequency (in units of $t_z$) for different temperatures at $\lambda=0.3$ and $c=0.01$.
In the upper inset, the width $\Gamma$ of the conductivity as a function of temperature at $\lambda=0.3$ and $c=0.01$. In the lower inset,
the conductivity in the limit of zero frequency as a function of $1/T$ at $\lambda=0.3$ and $c=0.01$.
The lattice size is $L=30^3$.}
\label{conduct1}
\end{figure}

Similar trends are obtained when the density of states is analyzed as function of $\lambda$. In Fig. \ref{density2}, the density of states as function of
$\lambda$ is reported at $T=0.3$. We notice that the tail is very small for $\lambda=0.4$, and it increases very rapidly as function of $\lambda$.
The tail energy $E_{tail}$ is plotted as function of $\lambda$ in the inset of Fig. \ref{density2}. We realize that $E_{tail}$ is a linear function
of $\lambda$. This behavior in $\lambda$ again can be useful for the interpretation of experimental data.
Moreover, it clarifies that, in the intermediate regime, $\lambda$ and $T$ play slightly different roles in contrast with
SCBA, where they are completely equivalent in enhancing the interaction of electrons with phonons.

From the analysis of Figs. \ref{density1} and \ref{density2}, it is clear that, in the regime of low carrier doping investigated
in this paper, the chemical potential shows a strong tendency to go towards the energy region of the tail with increasing the
temperature and the el-ph coupling. This is a robust feature in the intermediate regime. Therefore, this analysis
allows to reconcile the results provided by the ARPES data (band-like description) with the trend that charge carriers appear more localized
with increasing temperature. This behavior is very important for the discussion of optical and transport properties in the next section.

\section{Optical and transport properties}

As mentioned in the introduction, measurements of the optical and transport conductivity are important tools to investigate the properties of
organic semiconductors. For this reason we have calculated these quantities within the studied model. First, we have analyzed the weak el-ph
coupling regime within SCBA trying to understand the results of a band-like description. Then, we have considered the intermediate el-ph coupling regime.
We show the optical conductivity and mobility only along $z$ direction, since the corresponding quantities along the axis $x$ and $y$ can be
roughly reproduced taking into account the anisotropy factors $t_x/t_z$ and $t_y/t_z$, respectively.

\begin{figure}[htb]
\centering
\includegraphics[width=0.55\textwidth,angle=0]{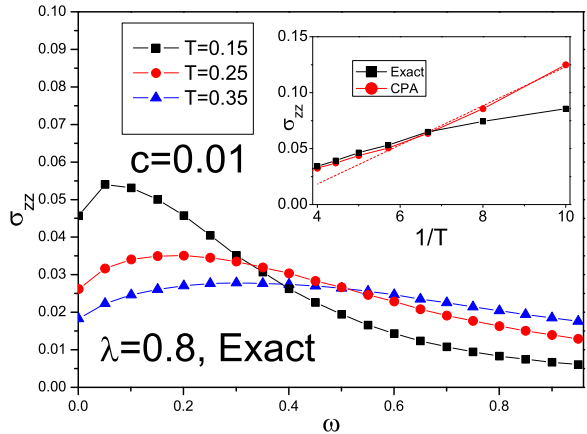}
\caption{(Color online) Optical conductivity $\sigma_{zz}$ within ED as a function of frequency (in units of $t_z$) for different temperatures at $\lambda=0.8$ and $c=0.01$.
In the inset, the conductivity $\sigma_{zz}$ in the limit of zero frequency within ED and CPA as a function of temperature at $\lambda=0.8$ and $c=0.01$.
Within ED, $L=10^3$ for conductivity at finite frequency, $12^3$ for conductivity in the limit of zero frequency. Within CPA, $L=26^3$.}
\label{conduct2}
\end{figure}

In Fig. \ref{conduct1}, we show the conductivity for different temperatures at $\lambda=0.3$ and $c=0.01$. For these values of parameters,
the optical conductivity follows a Drude behavior: $\sigma \propto \Gamma/(\omega^2+\Gamma^2)$,
with $\Gamma$ the width of the Drude peak. In the upper inset, $\Gamma$ is plotted as function of temperature: it follows a quadratic law.
Finally, in the lower inset, we report the conductivity in the limit of omega to zero. In the adiabatic model that we are studying,
the temperature behavior is proportional to the inverse of temperature. This is the description of band transport in our model.

The next step is to increase the el-ph coupling.
The results from ED simulations are necessary to the analysis since they clarify the effects of vertex corrections discarded by CPA. \cite{economou}
In Fig. \ref{conduct2}, the optical conductivity is reported for different temperatures at $\lambda=0.8$ and $c=0.01$. First, we notice
the reduction of conductivity in comparison with the data at $\lambda=0.3$. Moreover, the Drude peak does not longer represent the main spectral feature.
Indeed, the conductivity exhibits a clear peak at low frequencies $\omega < 0.6t_z$, whose intensity decreases with the temperature
moving slightly towards high energies. This behavior can be ascribed to the role played by the
vertex corrections into the calculation of conductivity since the peak does not appear in CPA. Moreover, the peak positions
do not depend on the broadening energy $\eta$ (used in the calculation of the conductivity), that we still choose equal to $3 \times 10^{-2} t_z$.
The result is of some interest since there are experimental evidences \cite{li,fisher} that, indeed, a peak is
present at energies around $62meV$  ($500cm^{-1}$) lower than any charge transfer process.\cite{petrenko}
This analysis confirms the findings of a previous paper for a related model. \cite{vittoriocheck}

In the inset of Fig. \ref{conduct2}, we report the comparison between the conductivity in the limit of zero frequency within ED and
that within CPA for different temperatures. It is clear that, in the low temperature regime, CPA overestimates the value of conductivity
pointing again towards the central role played by vertex corrections. The temperature behavior of $\sigma_{zz}$ is proportional to $1/T$,
like in SCBA. At high temperatures, the CPA and ED conductivities at zero frequency tend to assume close values
even if the finite frequency responses are quite different. Furthermore, the crossover from low to high temperature behavior
has a signature in the change of slope for both CPA and ED conductivities. The temperature behavior of
conductivity appears to be proportional to $1/T$ also at high temperature.

\begin{figure}[h]
\centering
\includegraphics[width=0.56\textwidth,angle=0]{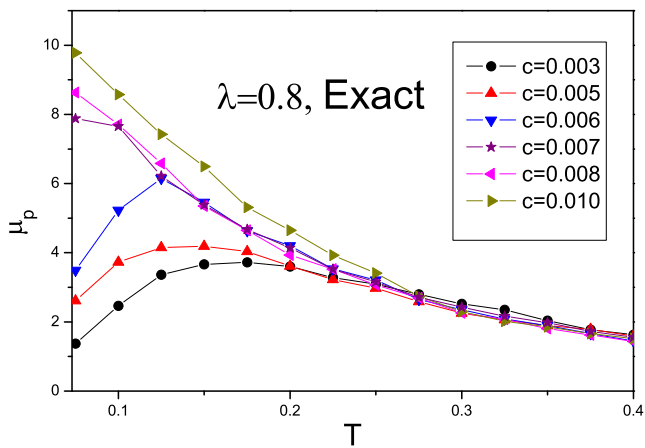}
\caption{(Color online) Mobility within ED as a function of temperature for different particle densities at $\lambda=0.8$.
The lattice size is $L=12^3$.}
\label{conduct3}
\end{figure}

Finally, we analyze the mobility $\mu_p$ along $z$ direction as function of temperature with changing the particle density.
As shown in Fig. \ref{conduct3}, the behavior is quite complex since it strongly depends also on the carrier density.
For densities around one per cent, as pointed out in the previous section, the chemical potential at low temperature is above the mobility edge.
Thus, the mobility is metallic-like. However, the behavior at low and high temperature is different although all the range is characterized
by a power law. Indeed, with increasing temperature, more and more states participate to the transport mechanism. However,
the inclusion of localized states and itinerant damped states is only able to decrease the magnitude of the mobility.

The situation is different for densities below one per cent. At those densities, the chemical potential is below the mobility edge, so that,
at low temperature, we expect that the mobility shows an insulating character. With increasing the temperature, as emphasized in the previous section,
the chemical potential is always in the tail regions, but more states at higher energies (of itinerant nature) get involved providing
the main contribution to mobility. Therefore, at intermediate temperatures, the mobility changes its character and exhibits a metallic power-law behavior. At higher temperatures, the mobilities look very similar for all the densities considered in this study.

 For temperatures $T \simeq \omega_0$, decreasing the densities, we observe that the chemical potential and, therefore, the most important states involved in the transport properties cross the temperature dependent mobility edge (see Fig.3). Translated in a polaronic framework, this would mean that the single particle sees a potential well due to el-ph coupling, hence, at very low densities, a polaron-like activated mechanism could come out. On the other hands, at higher temperatures, the scenario changes. The large broadening observed in the spectral function peaks at any density signals a new regime. As a result, the mobility decreases with the temperature acquiring a power-law behavior that do not depend significatively on density.

Finally, a comment on the behavior at very low temperatures ($T << \omega_0$) is in order. Indeed, in this regime, the adiabatic approach adopted in this work is no longer appropriate. As discussed in section II "The Model", at very low temperatures, quantum lattice fluctuations could play a role. On the other hand, at the temperatures $T \geq \omega_0 \simeq 120 K $ considered in this paper, our approach is fully consistent. This discussion clarifies in which sense  the comparison with experimental data in OFETs of rubrene and pentacene at very low densities and temperatures can be pursued. Of course, if the samples are not very pure, but contain impurities or dislocations, then other mechanisms, not included in the present work, should be taken into account. However, our analysis suggests that there is a narrow temperature window where an insulating behavior can be obtained without invoking the presence of impurities. In our view, the two different mechanisms could work together.

Finally, the mobilities shown in Fig. \ref{conduct3} are expressed in our natural units, that is in terms of $\mu_0 = e a^2 / \hbar$. Taking $a=7 \AA$, \cite{hannewald} one gets $\mu_0 \simeq 7 cm^2 /(V \cdot s)$. At room temperature $T=0.25 t_z$, the mobility is about $3 \mu_0 \simeq 21 cm^2 /(V \cdot s)$,
a value that recovers the right order of magnitude of experimental data in rubrene and pentacene. \cite{morpurgo} Moreover, the mobility results are consistent with those obtained within the picture of polaron formation. \cite{hannewald} Actually, the power-law behavior as function of temperature seems to be common to different mechanisms. In order to clarify the nature of the quasi-particles involved in the transport mechanism, it could be interesting to compare not only the mobility data, but also the optical and spectral properties of the present approach with those that could be calculated in the framework of the polaronic approaches. Actually, at the moment, these quantities in the polaronic framework are not available in the regime of parameters appropriate for organic semiconductors. Clearly, all the calculated properties have to be discussed in relation with experiments.

\section{Conclusions and Discussions}

Spectral, optical and transport properties have been calculated within the adiabatic three-dimensional anisotropic Holstein model
recently proposed for the study of organic semiconductors. Different approaches have been used: SCBA, CPA, and ED.

With increasing temperature, the width of the spectral functions $A(\vec{k},\omega)$ increases making the quasi-particle approximation less accurate. However, their peak positions are not strongly renormalized in comparison with the bare ones. These results have been discussed in relation with ARPES data that suggest a very simple quasi-particle energy dispersion. The marked width of the spectral functions is related to densities of states with a low energy exponential tail
increasing with temperature. The order of magnitude and the behavior of the tail energy as function of temperature and el-ph coupling have been analyzed
in connection with recent experiments.

The crossover between low and high temperature behavior seen in the spectral properties affects also the optical and transport properties.
In the weak el-ph coupling coherent regime, the optical conductivity follows a Drude behavior and the mobility can be interpreted in term of band transport.
In the intermediate el-ph coupling regime, at low temperature, vertex corrections introduced through ED give rise to a temperature dependent
low frequency peak in the conductivity, while, at higher temperatures, this peak broadens over large frequencies. This spectral feature agrees with the
results of a previous paper for a related model, \cite{vittoriocheck} and compares well with recent experimental results.\cite{li,petrenko,fisher}
Finally, for high temperatures, the mobility is characterized by a power-law behavior as function of $T$.
At low temperature, with decreasing the particle density, the mobility shows a transition from metallic to insulating behavior.
Results have been discussed in connection with available experimental data.

In the model investigated in this paper, the mobility scales as the inverse of temperature. In experimental data of rubrene and pentacene,
the mobility often decreases faster as function of temperature. In order to simulate more realistic systems,
the next step would be to consider Holstein-Peierls models, \cite{hannewald1,wang,ester} where, as mentioned in the Introduction, the lattice distortions affect also off-diagonal matrix elements of the electronic system. Actually, as shown also in recent papers by some of us, \cite{vittoriocheck,perronissh} the non-local el-ph couplings in organic semiconductors are not negligible, and they can affect the temperature dependence of carrier mobility.

Long-range el-ph couplings could be included within appropriate models \cite{perroni} in order to analyze the role of interface phonons
in OFETs. \cite{nature}  Finally, it would be interesting to investigate the role of quantum lattice fluctuations.
Their effects are small in the adiabatic limit, however, they could be important in the regime of very low temperatures.
Work in this direction is in progress.

\section{Acknowledgements}
This work was partially supported by University of Napoli "Federico II" under grant FARO.

\end{document}